\documentclass[12pt]{iopart}     
\usepackage{graphicx}

\begin{document}

\title{Cosmological Perturbations from a Group Theoretical Point of View}

\author{Istv\'an Szapudi$^1$, Viktor G.~Czinner$^{2,3}$}  
\address{$^1$Institute for Astronomy, University of Hawaii, 2680 Woodlawn Dr., 
Honolulu, HI 96815, USA}
\address{$^2$Department of Mathematics and Applied Mathematics, 
University of Cape Town, Rondebosch, 7701, South Africa}
\address{$^3$Department of Theoretical Physics, 
KFKI Research Institute for Particle and Nuclear Physics, 
H-1525 Budapest 114, P.O.~Box 49, Hungary}

\eads{\mailto{szapudi@ifa.hawaii.edu}, \mailto{czinner@rmki.kfki.hu}}

\begin{abstract}
We present a new approach to cosmological perturbations based on the
theory of Lie groups and their representations. After re-deriving the
standard covariant formalism from $SO(3)$ considerations, we provide 
a new expansion of the perturbed Friedmann-Lema\^{\i}tre-Robertson-Walker 
(FLRW) metric in terms of irreducible representations of the Lorentz group.
The resulting decomposition splits into (scalar, scalar), (scalar, vector) and 
(vector, vector) terms. These equations directly correspond to the standard Lifshitz 
classification of cosmological 
perturbations using scalar, vector and tensor modes which arise from the irreducible 
$SO(3)$ representation of the spatial part of the metric. While the Lorentz group 
basis matches the underlying local symmetries of the FLRW spacetime better than the 
$SO(3)$, the new equations do not provide further simplification compared to the 
standard cosmological perturbation theory. We conjecture that this is due to the 
fact that the $so(3,1)\sim su(2)\times su(2)$ Lorentz algebra has no pair of commuting 
generators commuting with any of the translation group generators.
\end{abstract}
\maketitle

\section{Introduction}\label{intro}

Linear cosmological perturbation theory plays a fundamental role in our understanding of 
the evolution of small inhomogeneities of the universe scaling from quantum fluctuations 
to large scale structure formation. In an early paper \cite{LK}, Lifshitz and Khalatnikov 
laid the groundwork for all subsequent research on the subject.  The cosmological 
perturbations were expressed in spatially flat gauge, concentrating
all fluctuations into the
spatial part of the metric. The perturbations were then classified according to their 
transformation properties in the background spacetime as scalar, vector and tensor modes, 
and it was shown that the different modes do not mix in the linear theory. This is the 
celebrated ''decomposition theorem''.

Based on the above results, almost twenty years later, Bardeen introduced a gauge invariant 
formalism \cite{B} which has since become the cornerstone of linear cosmological
perturbation theory. The formalism has been further developed and generalized in subsequent 
works (see e.g.~\cite{KS,MFB}).

In the gauge dependent approach of Lifshitz and Khalatnikov \cite{LK}, all gauge freedom  
is used to constrain the perturbations into the spatial domain. Due to the isotropy of the 
background FLRW spacetime, the spatial part of the perturbed metric is an $SO(3)$ tensor, and as 
such, it can be expanded into irreducible representations, an $l=0$ scalar (the trace of the 
spatial part of the metric) and an $l=2$ tensor (a traceless symmetric tensor). The latter 
has five independent components, corresponding to the $m = -2,-1,0,1,2$ eigenvalues of the 
rotation generator around an arbitrarily chosen direction. The classification of \cite{LK}
corresponds to $|m|$ of these representations, where $|m|=0,1,2$ belongs to the scalar-, vector- 
and tensor modes respectively (note the slightly different and confusing terminology from 
group theory). In other words, perturbations are classified according to how a mode responds 
to rotations around a chosen axis.

Interestingly in \cite{B}, Bardeen kept the same classification scheme even for gauge invariant 
perturbations. Once the whole metric is considered, it is locally transforming as 
an $SO(3,1)$, i.e.~a Lorentz tensor. The Lorentz algebra describes 
transformations of the time-time and time-space components of the metric under infinitesimal 
$so(3)$ rotations of space, a subalgebra of the full Lorentz algebra. It is easy to show that 
the time-time component of the metric transforms as a scalar with respect to spatial rotations, 
and the time-space part as a vector. As shown by Bardeen, the decomposition theorem still holds 
using the classification of the perturbations according to the $m$ eigenvalue 
(magnetic quantum number in physicist's terminology). 

The above considerations provide a motivation to investigate the mathematical theory of cosmological 
perturbations from a group theoretical point of view. In addition, in the literature it is customary 
to start from the result that the spatial dependence of linear perturbations in the Fourier-space can 
be completely represented by the solutions of the generalized Helmholtz equation. The mathematical 
background of this result is that the symmetry group of the solutions of the Helmholtz equation, 
the Euclidean group (see e.g.~\cite{M}), is precisely the group of geometrical symmetries of 
the spatial part of the (homogeneous and isotropic) FLRW metric. Thus the expansion of the spatial 
part of the perturbations into solutions of the Helmholtz equation is essentially a harmonic analysis 
on the Euclidean group as a Lie group. This observation provides further motivation to examine closely 
the group theoretical structure behind perturbation theory.

To the best of our knowledge, the theory of cosmological perturbations, in general, has not been 
considered to date from the point of view of Lie groups, and the principal goal of the present paper 
is to develop an approach to cosmological perturbations based on the theory of Lie groups and their 
representations. More specifically, we investigate the theory using irreducible representations derived 
from the more relevant Lorentz algebra, and not just its $so(3)$ subalgebra as usual. This is motivated 
by the observation that the spatially flat FLRW spacetime (which we only consider in this paper), i.e.~the 
zeroth order cosmological solution, is conformally flat, and by the locally Lorentzian nature of the metric 
tensor. Our hope is to elucidate the role that symmetries play in the decomposition of the 10 Einstein 
equations into independent subsets based on the decomposition theorem.

The plan of the paper is as follows. In section \ref{cov}, we present a quick overview of the standard 
covariant formalism; for further details please refer to \cite{B,KS,MFB} or, 
for a more informal introduction, \cite{Hu}. In section \ref{expansion}, first we reproduce the 
standard covariant formalism from $SO(3)$ considerations, and then we expand 
the perturbations using irreducible representations of the full Lorentz algebra. In section \ref{perteqs}, we 
present the perturbed field equations obtained from the Lorentz decomposition, and in section \ref{gauge}, 
we discuss gauge invariant quantities. Finally, in section \ref{concl}, we summarize 
our results and draw conclusions.

\section{Standard covariant formalism in a nutshell}\label{cov}

The background spacetime is described by the FLRW metric which we introduce in the conformal form
\begin{equation}
ds^{2} = g_{\mu\nu} dx^{\mu}dx^{\nu}= a^{2}(\eta) (-d\eta^{2} +\gamma_{ij}dx^{i} dx^{j} )\,,
\end{equation}
where $\eta$ is the conformal time variable and $\gamma_{ij}$ is the metric tensor for a 3-space of uniform
spatial curvature $K$.  Here, and throughout the paper, Greek indices run from $0$ to $3$ while 
Latin indices run from $1$ to $3$.

As mentioned earlier, perturbations in various quantities are classified according to
how they transform under spatial coordinate transformations in the
background spacetime as scalar-, vector- and tensor modes. The
homogeneity and isotropy of the background metric allows the
separation of the time dependence and the spatial dependence, with the
spatial dependence related to solutions of a generalized
Helmholtz-equation. Scalar-, vector- and tensor harmonics are solutions
of the scalar-
\[
 \nabla^2Q^{(0)}+k^2Q^{(0)}=0\,,
\]
vector-
\begin{equation}
\nabla^2Q^{(1)}_{i}+k^2Q^{(1)}_{i}=0\,,
\end{equation}
and tensor
\[
\nabla^2Q^{(2)}_{ij}+k^2Q^{(2)}_{ij}=0\
\]
Helmholtz-equations respectively, where $\nabla$ denotes the covariant
derivative with respect to the spatial metric $\gamma_{ij}$.

In a spatially flat ($K=0$) universe, the harmonics are essentially
plane waves:
\begin{eqnarray}
Q^{(0)}&=&\exp(i{\bf k}\cdot{\bf x})\,, \nonumber\\ 
Q_i^{(\pm 1)}&=&\frac{-i}{\sqrt{2}}({\bf e}_1 \pm i {\bf e}_2)_i \exp(i {\bf k}\cdot {\bf x})\,, \\ 
Q_{ij}^{(\pm 2)}&=& - \sqrt{\frac{3}{8}}({\bf e}_1 \pm i {\bf e}_2)_i ({\bf e}_1 \pm i {\bf e}_2)_j 
\exp(i {\bf k}\cdot {\bf x})\,, \nonumber
\end{eqnarray}
where ${\bf x}=(x_1,x_2,x_3)$ and ${\bf e}_{1}, {\bf e}_{2}$ are unit
vectors spanning the plane transverse to the wave vector ${\bf
  k}\equiv(k_1,k_2,k_3)$.  The vector modes represent divergence-free
(vorticity) vectors while the tensor modes are transverse and
traceless, and represent gravitational waves:
\begin{eqnarray}
\nabla^i Q_i^{(\pm 1)} = 0\,,\qquad \nabla^i Q_{ij}^{(\pm 2)} =
0\,,\qquad \gamma^{ij} Q_{ij}^{(\pm 2)} = 0\,.
\end{eqnarray}
The curl free vectors and the longitudinal components of tensors can
be obtained from covariant derivatives of the scalar- and vector modes
by
\begin{eqnarray}
Q_i^{(0)}&=&-k^{-1}\nabla_i Q^{(0)}\,, \nonumber
\\ Q_{ij}^{(0)}&=&(k^{-2} \nabla_i \nabla_j + {1 \over 3}
\gamma_{ij})Q^{(0)}\,, \\ Q_{ij}^{(\pm 1)}&=&-{1 \over 2k}[\nabla_i
  Q_j^{(\pm 1)}+\nabla_j Q_i^{(\pm 1)}] \,, \nonumber
\end{eqnarray}
where $k=|{\bf k}|$.

A completely general perturbation of the gravitational field can be
written as a linear combination of perturbations associated with
individual spatial harmonics defined above, and no coupling between the
different modes. Thus, a general perturbation to the FLRW metric can be
represented for example as \cite{B}
\begin{eqnarray}\label{metricpert}
g_{00}&=&-a^2(1+2A)\,, \nonumber\\ g_{0i}&=&-a^2B_i\,,
\\ g_{ij}&=&a^2(\gamma_{ij}+2H_L\gamma_{ij}+2H_{Tij})\,,
\nonumber\label{pertmetric}
\end{eqnarray}
with
\begin{eqnarray}
&A&\equiv A(x^k,\eta) - \mbox{scalar potential;}
  \nonumber\\ &B_i&\equiv B_i(x^k,\eta) - \mbox{vector
    shift;}\nonumber\\ &H_L&\equiv H_L(x^k,\eta) - \mbox{scalar
    perturbation to the spatial curvature;}\nonumber\\ &H_{Tij}&\equiv
  H_{Tij}(x^k,\eta) - \mbox{trace free distortion to the spatial
    metric.}\nonumber
\end{eqnarray}
For the $kth$ harmonics, the scalar-, vector- and tensor components of
the perturbed metric become
\begin{eqnarray}
A({\eta,\bf x})&=&A(\eta,k)Q^{(0)}({\bf x}) \,, \nonumber\\
H_L(\eta,{\bf x})&=&H_L(\eta,k)Q^{(0)}({\bf x}) \,, \nonumber\\ 
B_i (\eta,{\bf x})&=&\sum_{m=-1}^1 B^{(m)}(\eta,k)Q_i^{(m)}({\bf x})\,,\\ 
H_{Tij}({\eta,\bf x})&=&\sum_{m=-2}^{2}H_T^{(m)}(\eta,k)Q_{ij}^{(m)}({\bf x})\,.\nonumber
\end{eqnarray}

Let us now rewrite the perturbed metric (\ref{metricpert}) in the form
\begin{equation}
g_{\alpha\beta}=\mathring{g}_{\alpha\beta}+\delta g_{\alpha\beta}\,.
\end{equation}
It can be shown by simple calculations that according to the presented
expansion, the first order part of the metric 
\begin{eqnarray}\label{deltametric}
\delta g_{00}&=&-2a^2A\,, \nonumber\\ \delta g_{0i}&=&-a^2B_i\,,
\\ \delta g_{ij}&=&2a^2(H_L\gamma_{ij}+H_{Tij})\,, \nonumber
\end{eqnarray} 
with the choice of ${\bf k}=(0,0,k)$, can be decomposed as
\begin{equation}\label{gdec}
\delta g_{\alpha\beta}=\left(Q^{(0)}\sum_{j=1}^{10}
o_jO_j\right)_{\alpha\beta},
\end{equation}
where the $o_j$ amplitudes are
\begin{eqnarray}
o_1&=&2a^2A(\eta,k)\,,\quad\qquad
o_2=2a^2H_L(\eta,k)\,,\nonumber\\
o_3&=&\frac{a^2}{\sqrt{2}}B^{(-1)}(\eta,k)\,,\quad\ 
o_4=a^2B^{(0)}(\eta,k)\,,\quad\ \ 
o_5=\frac{a^2}{\sqrt{2}}B^{(+1)}(\eta,k)\,,\\
o_6&=&\sqrt{\frac{3}{2}}a^2H_T^{(-2)}(\eta,k)\,,\ 
o_7=\frac{a^2}{\sqrt{2}}H_T^{(-1)}(\eta,k)\,,\ 
o_8=\frac{2a^2}{3}H_T^{(0)}(\eta,k)\,,\nonumber\\
o_9&=&\frac{a^2}{\sqrt{2}}H_T^{(+1)}(\eta,k)\,,\quad
o_{10}=\sqrt{\frac{3}{2}}a^2H_T^{(+2)}(\eta,k)\,,\nonumber
\end{eqnarray}
and $O_j$ are the following 10, $4\times 4$, symmetric matrices:

\begin{eqnarray}
\!\!\!\!\!\!\!\!\!\!\!\!\!\!\!\!\!\!\!\!\!\!\!\!\!\!\!\!\!\!\!\!\!\!
O_1&=& \left(
\begin{array}{cccc}
-1 & 0 & 0 & 0\\ 0 & 0 & 0 & 0 \\ 0 & 0 & 0 & 0 \\ 0 & 0 & 0 & 0
\end{array}
\right);\qquad 
O_2= \left(
\begin{array}{cccc}
0 & 0 & 0 & 0\\ 0 & 1 & 0 & 0 \\ 0 & 0 & 1 & 0 \\ 0 & 0 & 0 & 1
\end{array}
\right);\nonumber\\ 
\!\!\!\!\!\!\!\!\!\!\!\!\!\!\!\!\!\!\!\!\!\!\!\!\!\!\!\!\!\!\!\!\!\!
O_3&=& \left(
\begin{array}{cccc}
0 & i & 1 & 0\\ i & 0 & 0 & 0 \\ 1 & 0 & 0 & 0 \\ 0 & 0 & 0 & 0
\end{array}
\right);\quad\quad \quad
O_4= \left(
\begin{array}{cccc}
0 & 0 & 0 & i\\ 0 & 0 & 0 & 0 \\ 0 & 0 & 0 & 0 \\ i & 0 & 0 & 0
\end{array}
\right);\quad 
O_5= \left(
\begin{array}{cccc}
0 & i & -1 & 0\\ i & 0 & 0 & 0 \\ -1 & 0 & 0 & 0 \\ 0 & 0 & 0 & 0
\end{array}
\right);\\ 
\!\!\!\!\!\!\!\!\!\!\!\!\!\!\!\!\!\!\!\!\!\!\!\!\!\!\!\!\!\!\!\!\!\!
O_6&=& \left(
\begin{array}{cccc}
0 & 0 & 0 & 0\\ 0 & -1 & i & 0 \\ 0 & i & 1 & 0 \\ 0 & 0 & 0 & 0
\end{array}
\right),\quad 
O_7= \left(
\begin{array}{cccc}
0 & 0 & 0 & 0\\ 0 & 0 & 0 & -1 \\ 0 & 0 & 0 & i \\ 0 & -1 & i & 0
\end{array}
\right);\quad 
O_8= \left(
\begin{array}{cccc}
0 & 0 & 0 & 0\\ 0 & 1 & 0 & 0 \\ 0 & 0 & 1 & 0 \\ 0 & 0 & 0 & -2
\end{array}
\right);\nonumber\\ 
\!\!\!\!\!\!\!\!\!\!\!\!\!\!\!\!\!\!\!\!\!\!\!\!\!\!\!\!\!\!\!\!\!\!
O_9&=& \left(
\begin{array}{cccc}
0 & 0 & 0 & 0\\ 0 & 0 & 0 & -1 \\ 0 & 0 & 0 & -i \\ 0 & -1 & -i & 0
\end{array}
\right); \qquad 
O_{10}= \left(
\begin{array}{cccc}
0 & 0 & 0 & 0\\ 0 &-1 & -i & 0 \\ 0 &-i & 1 & 0 \\ 0 & 0 & 0 & 0
\end{array}
\right).\nonumber
\end{eqnarray}\\
This decomposition provides a unique expansion of the metric perturbations
up to constant normalization factors.

The parametrization of the matter stress-energy tensor of a perfect fluid in the standard formalism is given by
\begin{eqnarray}
T^0_{\ 0}=-\rho-\delta\rho,\nonumber\\
T^0_{\ i}=(\rho+p)(v_i-B_i),\nonumber\\
T^i_{\ 0}=-(\rho+p)v_i,\nonumber\\
T^i_{\ j}=(p+\delta p)\delta^i_{\ j}+p\Pi^i_{\ j},\nonumber
\end{eqnarray}
where $\delta\rho$ is a scalar density perturbation, $v_i$ is a vector velocity perturbation
to the spacelike part of the zeroth order four velocity $u_{\alpha}=[a(\eta),0,0,0]$ of the fluid, and 
$\Pi_{ij}$ is a tensor anisotropic stress perturbation.

In the present paper we prefer to work with lower indices, thus for the perturbation of
the stress-energy tensor we use
\begin{equation}
\delta T _{\alpha\beta}=\mathring{g}_{\alpha\mu}\delta T^{\mu}_{\ \beta}-
\mathring{g}_{\alpha\nu}\delta g^{\nu\tau}\mathring{T}_{\tau\beta}.
\end{equation}
By taking into account the explicit form of $\mathring{T}_{\alpha\beta}$ as 
\begin{equation}
\mathring{T}_{\alpha\beta}=(\rho+p)u_{\alpha}u_{\beta}+p\mathring{g}_{\alpha\beta},
\end{equation}
with the given parametrization of the metric and the matter perturbations we get
\begin{eqnarray}
\delta T_{00}=a^2(\delta\rho+2\rho A),\nonumber\\
\delta T_{0i}=-a^2[(\rho+p)(v_i-B_i)+pB_i],\\
\delta T_{ij}=a^2[\delta p\delta_{ij}+p[\Pi_{ij}+2(H_L\gamma_{ij}+H_{Tij})]].\nonumber
\end{eqnarray} 

Now, in a completely analog fashion to the metric perturbations (given in (\ref{gdec})), we can decompose 
the perturbations of the stress-energy tensor as
\begin{equation}
\delta T_{\alpha\beta}=\left(Q^{(0)}\sum_{j=1}^{10}
T_jO_j\right)_{\alpha\beta} ,
\end{equation}
where, for the $kth$ harmonics, the scalar-, vector- and tensor components of
the perturbations become
\begin{eqnarray}
\delta\rho({\eta,\bf x})&=&\delta\rho(\eta,k)Q^{(0)}({\bf x}) \,, \nonumber\\
\delta p(\eta,{\bf x})&=&\delta p(\eta,k)Q^{(0)}({\bf x}) \,, \nonumber\\ 
v_i (\eta,{\bf x})&=&\sum_{m=-1}^1 v^{(m)}(\eta,k)Q_i^{(m)}({\bf x})\,,\\ 
\Pi_{ij}({\eta,\bf x})&=&\sum_{m=-2}^{2}\Pi^{(m)}(\eta,k)Q_{ij}^{(m)}({\bf x})\,,\nonumber
\end{eqnarray}
and the $T_j$ coefficients are 
\begin{eqnarray}
T_1&=&-a^2(\delta\rho+2\rho A)\,,\qquad\qquad\quad\ \ 
T_2=a^2(\delta p+2pH_L)\,,\nonumber\\
T_3&=&\frac{a^2}{\sqrt{2}}\left[(\rho+p)v^{(-1)}-\rho B^{(-1)}\right]\,,\quad
T_4=\frac{a^2}{\sqrt{2}}\left[(\rho+p)v^{(0)}-\rho B^{(0)}\right],\nonumber\\
T_5&=&\frac{a^2}{\sqrt{2}}\left[(\rho+p)v^{(+1)}-\rho B^{(+1)}\right]\,,\quad
T_6=\sqrt{\frac{3}{8}}a^2p\left(\Pi^{(-2)}+2H_T^{(-2)}\right),\\ 
T_7&=&\frac{a^2p}{2\sqrt{2}}\left(\Pi^{(-1)}+2H_T^{(-1)}\right)\,,\qquad\ \ \ 
T_8=\frac{a^2p}{3}\left(\Pi^{(0)}+2H_T^{(0)}\right),\nonumber\\
T_9&=&\frac{a^2p}{2\sqrt{2}}\left(\Pi^{(+1)}+2H_T^{(+1)}\right)\,,\qquad\quad
T_{10}=\sqrt{\frac{3}{8}}a^2p\left(\Pi^{(+2)}+2H_T^{(+2)}\right).\nonumber
\end{eqnarray}

With this parametrization it is straightforward, although tedious,
to derive the Einstein's field equations at linear order. The explicit 
forms of the scalar-, vector- and tensor equations are provided for example 
in the equations (17), (20) and (22) in \cite{Hu} respectively. We will use 
them as a reference hereafter, and also present them here in the spatially
flat ($K=0$) case for our latter convenience.
\\\\
\indent{\it Scalar equations}

\begin{eqnarray}\label{sHu}
&k^2[H_L+\frac{1}{3}H_T+\frac{\dot a}{a}(\frac{B}{k}-\frac{\dot H_T}{k^2})]
= 4\pi Ga^2 \left[\delta\rho+3\frac{\dot a}{a}(\rho+p)\frac{v-B}{k}\right] \,,\nonumber\\
&k^2(A+H_L+ {1 \over 3}H_T)+\left({d\over d\eta}+2{\dot a \over a}\right)(kB-\dot H_T) 
=-8\pi Ga^2p\Pi \,,\nonumber\\
&{\dot a \over a} A - \dot H_L -{1 \over 3}  \dot H_T =  4\pi G a^2 (\rho+p){v-B\over k} \,,\nonumber\\
&\left[2{\ddot a \over a}-2\left({\dot a \over a}\right)^2+{\dot a \over a}{d \over d\eta}
-{k^2 \over 3}\right]A-\left[{d\over d\eta}+{\dot a \over a}\right](\dot H_L+{kB\over 3})
=4\pi Ga^2(\delta p+{1\over 3}\delta\rho) . \nonumber
\end{eqnarray}
\\
\indent{\it Vector equations}

\begin{eqnarray}
&kB^{(\pm 1)}-\dot H_T^{(\pm 1)}= 16\pi G a^2(\rho+p){v^{(\pm 1)}-B^{(\pm 1)} \over k} \,,\nonumber\\
&\left[{d\over d\eta}+2{\dot a \over a}\right](kB^{(\pm 1)}-\dot H_T^{(\pm 1)})=-8\pi Ga^2p\Pi^{(\pm 1)} .\nonumber
\end{eqnarray}
\\
\indent{\it Tensor equations}
\\
\[
\left[{d^2\over d\eta^2}+2{\dot a\over a}{d\over d\eta}+k^2\right]H_T^{(\pm 2)}=8\pi Ga^2p\Pi^{(\pm 2)} .
\]

\section{Perturbation theory using irreducible representations}\label{expansion}

In the present section, we re-derive the perturbed field equations of the standard formalism
with more explicit reference to group theory. First we consider the expansion of the metric 
perturbations according to irreducible representation of the $SO(3)$ group, and show
that the corresponding equations are identical with the ones presented in the previous section. 
After this preparation, we consider the expansion according to an irreducible representation 
of $SO(3,1)$, the full Lorentz group. We explicitly compute a (trace-orthogonal) basis for 
the decomposition of a second grade $SO(3,1)$ tensor according to the $su(2)\times su(2)$ 
decomposition of the Lorentz algebra. We present the corresponding linear field equations 
in the next section.

\subsection{The $SO(3)$ case}
To specify our notation, the Lorentz algebra is defined by the
following commutation relations
\begin{equation}
[L_{\mu\nu},L_{\rho\sigma}] = i g_{\nu\rho}L_{\mu\sigma} -i
g_{\mu\rho}L_{\nu\sigma}-i g_{\nu\sigma}L_{\mu\rho}+ i
g_{\mu\sigma}L_{\nu\rho}\ .
\end{equation}
Note that we are using ''physicist'' convention, i.e.~Hermitian generators,
instead of the anti-Hermitian (or ''mathematician'') convention.
The spatial part of the algebra can be cast in a familiar form
\begin{equation}
  [J_j,J_k] = i \epsilon_{jkl}J_l\ ,
\end{equation}
where $J_j = \frac{1}{2} \epsilon_{jkl}L_{kl}$  is the Hodge dual of
the spatial generators. This subalgebra is equivalent to that of $su(2)$
or $so(3)$.  With the usual boost generators $K_i = L_{0i}$ we have the
following relations
\begin{eqnarray}
[K_j,K_k] = & -i \epsilon_{jkl}K_l\ , \quad \left[J_j,K_k\right] = & i \epsilon_{jkl}K_l\ .
\label{so31alg2}
\end{eqnarray}

For the FLRW solution in a spatially flat universe, the metric
tensor is conformally flat and its spatial part is conformally Euclidean. 
Moreover, the spatial part of the metric is a locally $SO(3)$ tensor.
Therefore it makes sense to expand the spatial part of the metric in 
terms of irreducible representations under $SO(3)$. The irreducible 
representations of a $3\times3$, symmetric matrix are the
trace ($l=0$, scalar), and a rank-2, traceless, symmetric tensor ($l=2$ mode).
Here $l$ corresponds to the eigenvalues of the Casimir operator (total 
angular momentum in physics terminology), and for $l=2$ there are five 
modes which correspond to the five degrees of freedom in a rank-2, 
traceless, symmetric matrix. These are usually parametrized according 
to the eigenvalue of the rotation generator in the direction of an 
arbitrarily chosen z-axis. The corresponding eigenvalues (magnetic 
quantum numbers in physics) are $m=-2,-1,0,1, 2$. These $V_{l,m}$ 
tensors are traditionally called ''polarization operators'', and we 
quote their explicit forms from \cite{Vars} as given below.

\begin{eqnarray}
\!\!\!\!\!\!\!\!\!\!\!\!\!\!\!\!\!\!\!\!\!\!\!\!\!\!\!\!\!\!\!\!\!\!
V_{2,2}&=&\frac{1}{2}\left(
\begin{array}{ccc}
-1 & -i & 0 \\
-i & 1 & 0 \\
0 & 0 & 0
\end{array}
\right),\quad
V_{2,1}=\frac{1}{2}\left(
\begin{array}{ccc}
0 & 0 & 1 \\
0 & 0 & i \\
1 & i & 0
\end{array}
\right),\quad
V_{2,0}=\frac{1}{\sqrt{6}}\left(
\begin{array}{ccc}
1 & 0 & 0 \\
0 & 1 & 0 \\
0 & 0 & -2
\end{array}
\right),\nonumber\\
\!\!\!\!\!\!\!\!\!\!\!\!\!\!\!\!\!\!\!\!\!\!\!\!\!\!\!\!\!\!\!\!\!\!\\
\!\!\!\!\!\!\!\!\!\!\!\!\!\!\!\!\!\!\!\!\!\!\!\!\!\!\!\!\!\!\!\!\!\!
V_{2,-1}&=&\frac{1}{2}\left(
\begin{array}{ccc}
0 & 0 & -1 \\
0 & 0 & i \\
-1 & i & 0
\end{array}
\right),\quad
V_{2,-2}=\frac{1}{2}\left(
\begin{array}{ccc}
-1 & i & 0 \\
i & 1 & 0 \\
0 & 0 & 0
\end{array}
\right),\quad
V_{0,0}=\frac{1}{\sqrt{3}}\left(
\begin{array}{ccc}
1 & 0 & 0 \\
0 &  1 & 0 \\
0 & 0 & 1
\end{array}
\right).\nonumber
\end{eqnarray}

These matrices yield a natural basis on which to expand $3\times3$
symmetric matrices, such as the spatial part of the metric tensor. 
Careful comparison reveals that the previously introduced $O_2$ matrix
corresponds to $V_{0,0}$, i.e.~the trace is an $SO(3)$ scalar, while 
$O_6\ldots O_{10}$ directly correspond to $V_{2,2}\ldots V_{2,-2}$, the $SO(3)$
tensor modes with the five possible values of $m$; the 
three dimensional representations were extended into four dimensions
with rows and columns of $0$'s. 

The time-time part of the metric tensor $g_{00}$ 
does not respond to $SO(3)$ rotations, i.e.~it is an $SO(3)$ scalar. In four 
dimensions it is represented with the matrix $O_1$. It can be shown 
(c.f.~eq.~(\ref{so31alg2})) that under 
rotations $g_{0i}$ is an $SO(3)$ vector. The spherical basis for a vector is
$g_{03}$, $g_{01}\pm i g_{02}$. This is represented with the matrices $O_4$, 
$O_5$ and $O_3$ respectively (up to a $-i$ factor introduced for convenience 
sake). These latter matrices contain zeros for their spatial part.

As it was noted in the introduction, and it also follows from the above argument, 
the standard parametrization of the metric (based on the solution of the 
Helmholtz equations) directly corresponds to an irreducible representation 
according to $SO(3)$, i.e.~spatial rotations. The resulting $10$ parameters 
repackage the $10$ degrees of freedom in a symmetric $4\times 4$ matrix according 
to irreducible representations of $SO(3)$.

These matrices are suitable parameters for an expansion motivated by 
the $SO(3)$ invariance of the spatial part of the FLRW solution at fixed time, and
the local $SO(3)$ invariance of space. The homogeneity (translation invariance) of the 
universe can be further utilized with Fourier transformation, i.e.~an expansion 
according to the irreducible representation of the translation generators with 
eigenvalue $k$. In particular, once a corresponding $k$-mode is fixed for the
transform, we can choose the orientation of our $z$-axis (global rotation 
invariance). This will zero out derivatives in the other two directions. 
Conventionally this idea is expressed using ''{\it transverse}'' and 
''{\it longitudinal} '' modes in real space. 

We thus recovered the perturbation theory, described in section
\ref{cov}, by expanding the perturbed metric according to the above
representation of the $SO(3)$ group, and Fourier transforming after
choosing the $z$-axis such that it is parallel with the $k$-mode in
question. This procedure exploited the maximal symmetry 
group of the Helmholtz equation, the Euclidean group, that is also 
the group of geometrical symmetries of the spatial part of the background 
solution. The corresponding perturbed field equations are 
identical to the ones of the standard covariant formalism. Using
computer algebra, we explicitly recovered the equations of \cite{Hu}
based on the matrix formulation presented above.

\subsection{The $SO(3,1)$ case}
Let us now take into consideration the larger local symmetry of the metric tensor, the $SO(3,1)$ group, 
for constructing a perturbation theory according to irreducible representations. This is also motivated by 
the conformal flatness of the spatially flat FLRW solution. The calculation is exactly analogous to the 
exposition of the $SO(3)$ perturbation theory above. A popular parametrization of the $so(3,1)$ 
algebra splits into two commuting $su(2)$ algebras. Then for each $su(2)$ 
we can use the familiar theory of irreducible representations.

To find the irreducible representations, we used a spinor formalism: in each $su(2)$'s the irreducible 
representations are found by the standard way of symmetrizing spinor expressions. The direct product of 
the two commuting irreducible representations form a representation of $so(3,1)$. For constructing spinor 
representations we used Infeld-van der Waerden symbols (see e.g.~\cite{ODonnell})
\begin{eqnarray}
\sigma^0_{AB^\prime} &=& \frac{1}{\sqrt{2}}
\left( \begin{array}{cc}
1 & 0  \\
0 & 1 \\
\end{array} \right), \quad 
\sigma^1_{AB^\prime} = \frac{1}{\sqrt{2}}
\left( \begin{array}{cc}
0 & 1  \\
1 & 0 \\
\end{array} \right),\nonumber\\
\\
\sigma^2_{AB^\prime} &=& \frac{1}{\sqrt{2}}
\left( \begin{array}{cc}
0 & i  \\
-i & 0 \\
\end{array} \right), \quad 
\sigma^3_{AB^\prime} = \frac{1}{\sqrt{2}}
\left( \begin{array}{cc}
1 & 0  \\
0 & -1 \\
\end{array} \right)\nonumber
\end{eqnarray}
to map $g_{\mu\nu}$ into $g_{A A^\prime B B^\prime}$, where primed and
unprimed indices live in the two commuting $su(2)$'s. The symmetric
$4\times 4$ metric tensor is equivalent to a direct product of two
vector representations for each of the $su(2)$'s. One can find matrices
analogous to $O_j$ by symmetrizing in the primed and unprimed indices.
After explicit symmetrization we have one scalar mode: the $4D$
trace ($S$ in (\ref{Smatrixok})). The traceless $4\times 4$ tensor 
is a direct product of two vector modes, each of which can be classified 
according to the respective $m$ eigenvalue, corresponding to the generator 
according to an arbitrarily chosen $z$-axis in each representation. 
The pair $(m,m^\prime)$, both take three possible values, $-1,0,1$, giving 
9 possibilities. These 1+9 modes cover the 10 degrees of freedom in $4\times 4$ 
symmetric tensors. We present the corresponding $S$ and $S_{m,m^\prime}$ matrices 
explicitly:

\begin{eqnarray}\label{Smatrixok}
\!\!\!\!\!\!\!\!\!\!\!\!\!\!\!\!\!\!\!\!\!\!\!\!\!\!\!\!\!\!\!\!\!\!
S_{1,-1}&=&
\left(
\begin{array}{cccc}
1 & 0 & 0 & 1\\
0 & 0 & 0 & 0 \\
0 & 0 &  0 & 0 \\
1 & 0 & 0 & 1
\end{array}
\right);\qquad\quad\ 
S_{1,0}=
\left(
\begin{array}{cccc}
0 & 1 & i & 0\\
1 & 0 & 0 & 1 \\
i & 0 & 0 & i \\
0 & 1 & i & 0
\end{array}
\right);\nonumber\\
\!\!\!\!\!\!\!\!\!\!\!\!\!\!\!\!\!\!\!\!\!\!\!\!\!\!\!\!\!\!\!\!\!\!
S_{1,1}&=&
\left(
\begin{array}{cccc}
0 & 0 & 0 & 0\\
0 & 1 & i & 0 \\
0 & i & -1 & 0 \\
0 & 0 & 0 & 0
\end{array}
\right);\qquad
S_{0,-1}=
\left(
\begin{array}{cccc}
0 & 1 & -i & 0\\
1 & 0 & 0 & 1 \\
-i & 0 & 0 & -i \\
0 & 1 & -i & 0
\end{array}
\right);\nonumber\\
\!\!\!\!\!\!\!\!\!\!\!\!\!\!\!\!\!\!\!\!\!\!\!\!\!\!\!\!\!\!\!\!\!\!
S_{0,0}&=&
\left(
\begin{array}{cccc}
1 & 0 & 0 & 0\\
0 & 1 & 0 & 0 \\
0 & 0 & 1 & 0 \\
0 & 0 & 0 & -1
\end{array}
\right);
\qquad\ \ 
S_{0,1}=
\left(
\begin{array}{cccc}
0 & 1 & i & 0\\
1 & 0 & 0 & -1 \\
i & 0 &  0 & -i \\
0 & -1 & -i & 0
\end{array}
\right);\\ 
\!\!\!\!\!\!\!\!\!\!\!\!\!\!\!\!\!\!\!\!\!\!\!\!\!\!\!\!\!\!\!\!\!\!
S_{-1,-1}&=&
\left(
\begin{array}{cccc}
0 & 0 & 0 & 0\\
0 & 1 & -i & 0 \\
0 & -i & -1 & 0 \\
0 & 0 & 0 & 0
\end{array}
\right);
\quad\ \ 
S_{-1,0}=
\left(
\begin{array}{cccc}
0 & 1 & -i & 0\\
1 & 0 & 0 & -1 \\
-i & 0 & 0 & i \\
0 & -1 & i & 0
\end{array}
\right);\nonumber\\ 
\!\!\!\!\!\!\!\!\!\!\!\!\!\!\!\!\!\!\!\!\!\!\!\!\!\!\!\!\!\!\!\!\!\!
S_{-1,1}&=&
\left(
\begin{array}{cccc}
1 & 0 & 0 & -1\\
0 & 0 & 0 & 0 \\
0 & 0 & 0 & 0 \\
-1 & 0 & 0 & 1
\end{array}
\right);
\qquad\quad
S=
\left(
\begin{array}{cccc}
-1 & 0 & 0 & 0\\
0 & 1 & 0 & 0 \\
0 & 0 & 1 & 0 \\
0 & 0 & 0 & 1
\end{array}
\right).\nonumber
\end{eqnarray}

Similar to the $SO(3)$ case, we can decompose the perturbed metric according to the
above $su(2)\times su(2)$ representation as 
\begin{equation}\label{sdg}
\delta g_{\alpha\beta}=a^2\left[sS+\!\!\!\!\!\!\sum_{m,m'=-1}^{1}\!\!\!\!\!\!s_{m,m'}S_{m,m'}\right]_{\alpha\beta}\ , 
\end{equation}
where
\[
s_{m,m'}=s_{m,m'}(x^{\alpha}),\qquad s=s(x^{\alpha}) ,
\]
and the stress-energy tensor perturbation as
\begin{eqnarray}
\delta T_{00}=-a^2\left[(t+\rho s)S+\!\!\!\!\!\!\sum_{m,m'=-1}^{1}\!\!\!\!\!\!(t_{m,m'}
-\rho s_{m,m'})S_{m,m'}\right]_{00}\ ,\nonumber\\
\delta T_{0i}=a^2\left[(-t+ps)S+\!\!\!\!\!\!\sum_{m,m'=-1}^{1}\!\!\!\!\!\!(-t_{m,m'}
+ps_{m,m'})S_{m,m'}\right]_{0i}\ ,\\
\delta T_{ij}=a^2\left[(t+\rho s)S+\!\!\!\!\!\!\sum_{m,m'=-1}^{1}\!\!\!\!\!\!(t_{m,m'}
+ps_{m,m'})S_{m,m'}\right]_{ij}\ ,\nonumber 
\end{eqnarray}
with
\[
t_{m,m'}=t_{m,m'}(x^{\alpha}),\qquad t=t(x^{\alpha}) .
\]

Once we expand Einstein's equation into the parameters corresponding
to the above matrices, fix a $k$-mode for Fourier transforming, and
choose the $z$-axis parallel with this mode, we obtain the perturbation 
equations.

\section{Perturbation equations}\label{perteqs}

The decomposition of the perturbations according to the $S_{m,m'}$ 
and $S$ matrices results three independent groups of equations. A simple relation 
maps our decomposition to that of the traditional one, namely $m+m' \rightarrow m$, 
where $(m,m')$ corresponds to the $su(2)\times su(2)$, and $m$ corresponds to 
the traditional $so(3)$ scalar-vector-tensor decomposition. For example $(0,1)$ or 
$(-1,0)$ correspond to vector modes, etc. In addition, our parameter $s$ also 
corresponds to a scalar mode in scalar-vector-tensor terms.

Accordingly, the first group consists of two independent wave equations for $s_{1,1}$ 
and $s_{-1,-1}$ as
\begin{eqnarray}
&\ddot s_{1,1}+\frac{2\dot a}{a}\dot s_{1,1} +k^2s_{1,1}=16\pi G a^2 t_{1,1}\ ,\\
&\ddot s_{-1,-1}+\frac{2\dot a}{a}\dot s_{-1,-1} +k^2s_{-1,-1}=16\pi G a^2 t_{-1,-1}\ ,
\end{eqnarray}
where, and throughout the paper, an overdot denotes the derivative with respect to the conformal 
time $\eta$. The above equations are then clearly identical with the tensor equations of the standard 
formalism provided in Sec.2, and thus the $s_{1,1}$ and $s_{-1,-1}$ 
amplitudes correspond to gravitational waves while the $t_{1,1}$ and $t_{-1,-1}$
amplitudes correspond to anisotropic stress perturbations.

The second group of equations consists of two decoupled systems, one for $s_{1,0}$ and $s_{0,1}$,
and another for $s_{0,-1}$ and $s_{-1,0}$. The two systems are identical. 
\begin{eqnarray}\label{veqs1}
\!\!\!\!\!\!\!\!\!\!\!\!\!\!\!\!\!\!\!\!\!\!\!\!\!\!\!\!\!\!\!\!\!
&k(s_{0,1}+s_{1,0})-i(\dot s_{0,1}-\dot s_{1,0})=
\frac{16\pi G a^2}{k}[t_{0,1}+t_{1,0}-(\rho+p)(s_{0,1}+s_{1,0})],\\
\!\!\!\!\!\!\!\!\!\!\!\!\!\!\!\!\!\!\!\!\!\!\!\!\!\!\!\!\!\!\!\!\!\!\!\!\!\!\!\!\!\!\!\!\!
&k(s_{-1,0}+s_{0,-1})-i(\dot s_{-1,0}-\dot s_{0,-1})=
\frac{16\pi G a^2}{k}[t_{-1,0}+t_{0,-1}-(\rho+p)(s_{-1,0}+s_{0,-1})],\\
\!\!\!\!\!\!\!\!\!\!\!\!\!\!\!\!\!\!\!\!\!\!\!\!\!\!\!\!\!\!\!\!\!
&\left[\frac{d}{d\eta}+\frac{2\dot a}{a}\right](ik(s_{0,1}+s_{1,0})+(\dot s_{0,1}-\dot s_{1,0})=
16\pi G a^2(t_{0,1}-t_{1,0}),\\
\!\!\!\!\!\!\!\!\!\!\!\!\!\!\!\!\!\!\!\!\!\!\!\!\!\!\!\!\!\!\!\!\!
&\left[\frac{d}{d\eta}+\frac{2\dot a}{a}\right](ik(s_{-1,0}+s_{0,-1})+(\dot s_{-1,0}-\dot s_{0,-1})=
16\pi G a^2(t_{-1,0}-t_{0,-1}).
\label{veqs4}
\end{eqnarray}

Thus, the $s_{1,0}$, $s_{0,1}$, $t_{1,0}$, $t_{0,1}$ and $s_{-1,0}$, $s_{0,-1}$, $t_{-1,0}$, $t_{0,-1}$ degrees of 
freedom in the $su(2)\times su(2)$ representation correspond to the vector amplitudes of the $so(3)$ 
representation.

The remaining group of equations is the following coupled system of the $s_{1,-1}$, $s_{0,0}$, $s_{-1,1}$, $s$ and 
$t_{1,-1}$, $t_{0,0}$, $t_{-1,1}$, $t$ amplitudes,
{\small
\begin{eqnarray}
\!\!\!\!\!\!\!\!\!\!\!\!\!\!\!\!\!\!\!\!\!\!\!\!\!\!\!\!\!\!\!\!\!\!\!\!\!
&\frac{3\dot a^2}{a^2}(s_{0,0}+s_{1,-1}+s_{-1,1}-s)+
\frac{\dot a}{a}\left[\dot s_{0,0}+\dot s_{1,-1}+\dot s_{-1,1}+3\dot s-2ik(s_{1,-1}-s_{-1,1})\right]\\
\!\!\!\!\!\!\!\!\!\!\!\!\!\!\!\!\!\!\!\!\!\!\!\!\!\!\!\!\!\!\!\!\!\!\!\!\!
&+k^2(s_{0,0}+s)=-8\pi Ga^2(t_{0,0}+t_{1,-1}+t_{-1,1}+t+2\rho s),\nonumber\\
\!\!\!\!\!\!\!\!\!\!\!\!\!\!\!\!\!\!\!\!\!\!\!\!\!\!\!\!\!\!\!\!\!\!\!\!\!
&\left[\frac{d}{d\eta}+\frac{2\dot a}{a}\right](\dot s_{1,-1}+\dot s_{-1,1}-2 \dot s_{0,0}+
2ik(s_{-1,1}-s_{1,-1}))-k^2(s_{1,-1}+s_{-1,1}-2s)\\
\!\!\!\!\!\!\!\!\!\!\!\!\!\!\!\!\!\!\!\!\!\!\!\!\!\!\!\!\!\!\!\!\!\!\!\!\!
&=16\pi Ga^2(t_{1,-1}+t_{-1,1}-2t_{00}),\nonumber\\
\!\!\!\!\!\!\!\!\!\!\!\!\!\!\!\!\!\!\!\!\!\!\!\!\!\!\!\!\!\!\!\!\!\!\!\!\!
&i\left[\frac{\dot a}{a}(s-s_{1,-1}-s_{-1,1}-s_{0,0})-\dot s_{0,0}-\dot s\right]=
\frac{8\pi Ga^2}{k}(t_{1,-1}-t_{-1,1}-(\rho+p)(s_{1,-1}-s_{-1,1})),\\
\!\!\!\!\!\!\!\!\!\!\!\!\!\!\!\!\!\!\!\!\!\!\!\!\!\!\!\!\!\!\!\!\!\!\!\!\!
&\left[6\left(\frac{\dot a^2}{a^2}-\frac{\ddot a}{a}\right)-\frac{d}{d\eta}+k^2\right](s_{0,0}+s_{1,-1}+s_{-1,1}-s)-
4\left[\frac{\dot a}{a}(\dot s_{0,0}+\dot s_{1,-1}+\dot s_{-1,1})+\ddot s\right]+\\
\!\!\!\!\!\!\!\!\!\!\!\!\!\!\!\!\!\!\!\!\!\!\!\!\!\!\!\!\!\!\!\!\!\!\!\!\!
&2ik\left[\left(\frac{\dot a}{a}+\frac{d}{d\eta}\right)(s_{1,-1}-s_{-1,1})\right]=16\pi Ga^2(t-\rho s).\nonumber
\end{eqnarray}}
This system is clearly equivalent to the scalar equations of the $so(3)$ representation. 

In general, one can find the following one-to-one correspondence between the standard $so(3)$ 
and the $su(2)\times su(2)$ parametrization of the perturbation amplitudes  
\begin{eqnarray}
\!\!\!\!\!\!\!\!\!\!\!\!\!\!\!\!\!\!\!\!\!\!\!\!\!\!\!\!
s_{1,1} = -\sqrt{\frac{3}{2}}H_T^{(+2)},\qquad\qquad \qquad t_{1,1} = -\sqrt{\frac{3}{8}}p\Pi^{(+2)},\nonumber \\
\!\!\!\!\!\!\!\!\!\!\!\!\!\!\!\!\!\!\!\!\!\!\!\!\!\!\!\!
s_{-1,-1} = -\sqrt{\frac{3}{2}}H_T^{(-2)},\qquad \qquad \quad t_{-1,-1} = -\sqrt{\frac{3}{8}}p\Pi^{(-2)},\nonumber \\
\!\!\!\!\!\!\!\!\!\!\!\!\!\!\!\!\!\!\!\!\!\!\!\!\!\!\!\!
s_{\pm1,0} = \frac{iB^{(\pm1)}\mp H_T^{(\pm1)}}{2\sqrt{2}},\qquad \qquad
t_{\pm1,0} = \frac{1}{2\sqrt{2}}\left((\rho+p)iv^{(\pm1)}\mp \frac{p\Pi^{(\pm1)}}{2}\right),\nonumber\\ 
\!\!\!\!\!\!\!\!\!\!\!\!\!\!\!\!\!\!\!\!\!\!\!\!\!\!\!\!
s_{0,\pm1} = \frac{iB^{(\pm1)}\pm H_T^{(\pm1)}}{2\sqrt{2}},\qquad \qquad
t_{\pm1,0} = \frac{1}{2\sqrt{2}}\left((\rho+p)iv^{(\pm1)}\pm \frac{p\Pi^{(\pm1)}}{2}\right),\\ 
\!\!\!\!\!\!\!\!\!\!\!\!\!\!\!\!\!\!\!\!\!\!\!\!\!\!\!\!
s_{1,-1} = \frac{H_L-A+iB^{(0)}}{2}-\frac{H_T^{(0)}}{3},\quad 
t_{1,-1} =-\frac{1}{2}\left(\delta p+\delta\rho+\frac{p\Pi^{(0)}}{3}-(\rho+p)iv^{(0)}\right)\nonumber\\ 
\!\!\!\!\!\!\!\!\!\!\!\!\!\!\!\!\!\!\!\!\!\!\!\!\!\!\!\!
s_{-1,1} = \frac{H_L-A-iB^{(0)}}{2}-\frac{H_T^{(0)}}{3},\quad 
t_{-1,1} =-\frac{1}{2}\left(\delta p+\delta\rho+\frac{p\Pi^{(0)}}{3}+(\rho+p)iv^{(0)}\right)\nonumber\\
\!\!\!\!\!\!\!\!\!\!\!\!\!\!\!\!\!\!\!\!\!\!\!\!\!\!\!\!
s_{0,0}=\frac{H_L-A}{2}+\frac{2}{3}H_T^{(0)},\qquad \qquad t_{0,0}=\frac{p\Pi^{(0)}}{3}-\frac{\delta p+\delta\rho}{2}\nonumber\\
\!\!\!\!\!\!\!\!\!\!\!\!\!\!\!\!\!\!\!\!\!\!\!\!\!\!\!\!
s=\frac{A+3H_L}{2},\qquad \qquad\qquad \qquad t=\frac{3\delta p+\delta\rho}{2}\,\nonumber.
\end{eqnarray}

Despite the property that the Lorentz group matches the underlying local symmetries 
of the FLRW spacetime better than $SO(3)$, no further simplifications arise in the field equations 
compared to the standard decomposition theorem. We could not further decouple the above system. 
The new equations based on the irreducible representation of the $SO(3,1)$ group 
through an $su(2)\times su(2)$ decomposition of the Lorentz algebra appear to be equivalent to the 
$SO(3)$ approach.
 
\section{Gauge invariant quantities}\label{gauge}

One of the key features of Bardeen's formalism is the systematic 
determination of the gauge invariant variables. It is therefore important to show how these quantities
can be obtained under the $su(2)\times su(2)$ representation. In the present section we derive the transformation
properties of the metric perturbation amplitudes $s$ and $s_{m,m'}$, and compute the corresponding gauge invariant
metric quantities. For the matter perturbations the computations are completely analogous.

The most general gauge transformation is induced by the coordinate transformation  
\begin{equation}
\bar x^{\alpha}=x^{\alpha}+\xi^{\alpha}(x^{\beta}),
\end{equation}
where $\xi^{\alpha}$ is of the same order as $s$ and $s_{m,m'}$.

The changes in the metric tensor can be computed from the transformation law
\begin{equation}
\bar g_{\alpha\beta}(\bar x^{\gamma})=
\frac{\partial x^{\mu}}{\partial \bar x^{\alpha}}\frac{\partial x^{\nu}}{\partial \bar x^{\beta}}g_{\mu\nu}(x^{\kappa}),
\end{equation}
and the scale factors in $\bar g_{\alpha\beta}$ and $g_{\alpha\beta}$ at linear order are related by
\begin{equation}
a(\bar\eta)=a(\eta)\left[1+\frac{\dot a}{a}\xi^0\right]. 
\end{equation}
For the linear metric perturbations one can obtain the relation
\begin{equation}
\delta\bar g_{\alpha\beta}(\bar x^{\gamma})=\delta g_{\alpha\beta}(x^{\kappa})
-a^2(\eta)\left[\frac{2\dot a}{a}\xi^0\eta_{\alpha\beta} 
+ \eta_{\alpha\mu}\xi^{\mu}_{\ ,\beta}+\eta_{\nu\beta}\xi^{\nu}_{\ ,\alpha}\right],
\end{equation}
where $\eta_{\mu\nu}$ is the Minkowski metric tensor, and a comma denotes partial derivative. 
After inserting the explicit $su(2)\times su(2)$ decomposed forms of $\delta g_{\alpha\beta}$ 
and $\delta\bar g_{\alpha\beta}$ using (\ref{sdg}) we get
\begin{equation}
\!\!\!\!\!\!\!\!\!\!\!\!\!\!\!\!\!\!\!\!\!\!\!\!\!\!\!\!\!\!\!
\left[\bar sS+\!\!\!\!\!\!\sum_{m,m'=-1}^{1}\!\!\!\!\!\!\bar s_{m,m'}S_{m,m'}\right]_{\alpha\beta}=
\left[sS+\!\!\!\!\!\!\sum_{m,m'=-1}^{1}\!\!\!\!\!\!s_{m,m'}S_{m,m'}\right]_{\alpha\beta}\!\!\!\!\!\!-
\frac{2\dot a}{a}\xi^0\eta_{\alpha\beta}-\eta_{\alpha\mu}\xi^{\mu}_{\ ,\beta}-\eta_{\nu\beta}\xi^{\nu}_{\ ,\alpha}.
\end{equation}

The final form of the change in the amplitudes is then given by
\begin{eqnarray}
\bar s_{1,1} &=& s_{1,1},\\  
\bar s_{-1,-1} &=& s_{-1,-1},\\
\bar s_{1,0} &=& s_{1,0}+\frac{1}{4}\left(-\dot\xi^1+i\dot\xi^2-ik\xi^1-k\xi^2\right),\\
\bar s_{-1,0} &=& s_{-1,0}+\frac{1}{4}\left(-\dot\xi^1-i\dot\xi^2+ik\xi^1-k\xi^2\right),\\
\bar s_{0,1} &=& s_{0,1}+\frac{1}{4}\left(-\dot\xi^1+i\dot\xi^2+ik\xi^1+k\xi^2\right),\\
\bar s_{0,-1} &=& s_{0,-1}+\frac{1}{4}\left(-\dot\xi^1-i\dot\xi^2-ik\xi^1+k\xi^2\right),\\  
\bar s_{1,-1} &=& s_{1,-1}+\frac{1}{2}\left(\dot\xi^0-\dot\xi^3+ik\xi^0-ik\xi^3\right),\\
\bar s_{-1,1} &=& s_{-1,1}+\frac{1}{2}\left(\dot\xi^0+\dot\xi^3-ik\xi^0-ik\xi^3\right),\\
\bar s_{0,0} &=& s_{0,0}+\frac{1}{2}\left(\dot\xi^0+ik\xi^3\right),\\
\bar s &=& s+\frac{1}{2}\left(-\dot\xi^0-\frac{2\dot a}{a}\xi^0-ik\xi^3\right).
\end{eqnarray}

The $s_{1,1}$ and $s_{-1,-1}$ amplitudes are gauge invariant, just as we expected since they
correspond to the tensor modes in the standard formalism. Similarly to the field equations, the
remaining amplitudes decouple into two groups. The amplitudes belonging to the vector modes
in the standard formalism depend only on $\xi^1$ and $\xi^2$, while the amplitudes belonging 
to the scalar modes depend only on $\xi^0$ and $\xi^3$.

Consequently, from the first group, one can construct two gauge invariant variables
\begin{eqnarray}
\Psi^{(+1)}&=&i(s_{1,0}+s_{0,1})+\frac{\dot s_{0,1}-\dot s_{1,0}}{k}\ ,\\
\Psi^{(-1)}&=&i(s_{-1,0}+s_{0,-1})+\frac{\dot s_{-1,0}-\dot s_{0,-1}}{k}\ ,
\end{eqnarray}
which coincide with Bardeen's gauge invariant vector amplitudes. 

From the second group one can again construct two gauge invariant quantities
\begin{eqnarray}
\Phi_1&=&s+s_{0,0}+\frac{\dot a}{a}\left[\frac{2i}{k}(s_{-1,1}-s_{1,-1})
+\frac{1}{k^2}(\dot s_{-1,1}+\dot s_{1,-1}-2\dot s_{0,0})\right]\ ,\\ 
\Phi_2&=&s-s_{0,0}-s_{1,-1}-s_{-1,1}+\frac{2i}{k}\left[\dot s_{-1,1}-\dot s_{1,-1}
+\frac{\dot a}{a}(s_{-1,1}-s_{1,-1})\right]\nonumber\\
&-&\frac{1}{k^2}\left[2\ddot s_{0,0}-\ddot s_{1,-1}-\ddot s_{-1,1} 
+\frac{\dot a}{a}(2\dot s_{0,0}-\dot s_{1,-1}-\dot s_{-1,1})\right]\ ,
\end{eqnarray}
which, as one can check, are related to Bardeen's scalar variables as
\begin{equation}
\Phi_1\equiv 2\Phi_{H} \qquad\mbox{and}\qquad \Phi_2\equiv 2\Phi_{A}.
\end{equation}

As it is well known from the standard $so(3)$ theory (see e.g.~\cite{KS}), 
any gauge invariant variable that can be
constructed from $s_{1,0}$, $s_{-1,0}$, $s_{0,1}$ and $s_{0,-1}$ and their
time derivatives can be written as a linear combination of $\Psi^{(+1)}$ 
and $\Psi^{(-1)}$ and their time derivatives with coefficients of arbitrary
functions of time. The same is true for the amplitudes $s_{1,-1}$, $s_{-1,1}$, 
$s_{0,0}$ and $s$ with the variables $\Phi_1$ and $\Phi_2$. In addition,
corresponding pair of gauge invariant variables can be obtained from the 
perturbations of the stress-energy tensor to form the complete, closed set 
of gauge invariant variables (up to linear transformations).

\section{Summary and Conclusions}\label{concl}

We have developed an approach to linear cosmological perturbation theory based
on Lie groups and their representations. After a short 
overview on the standard formalism, based on the decomposition by the solutions of the 
generalized Helmholtz equation, we explicitly reproduced the theory from the
irreducible representations of the $SO(3)$ group, and utilizing the translation symmetry 
by Fourier transforming. This procedure completely exploited the maximal symmetry group 
of the Helmholtz equation (i.e.~the Euclidean group), the group of geometrical 
symmetries of the spatial part of the background FLRW solution. 

As an application of our new approach, we expanded the perturbations 
according to irreducible representations of the Lorentz algebra through the popular
$su(2)\times su(2)$ decomposition. This expansion was motivated by the fact that the 
spatially flat FLRW cosmological metric is conformally flat, and that the local symmetry group 
of the metric is the Lorentz group, of which $SO(3)$ is only a subgroup. Our hope was to elucidate 
the role that local symmetries play in the decomposition of the perturbed field equations, 
and check whether it is possible to obtain a simpler system of equations than that of the 
standard formalism. 

As a result we presented a new decomposition by (scalar, scalar), (scalar, vector), 
and (vector, vector) terms. We showed that these can be mapped directly  into the standard scalar-vector-tensor
classification. Furthermore, we showed that the resulting equations are simply related to 
the equations of \cite{Hu}. The decomposition according to the larger Lorentz group did not 
produce a finer split of the perturbed field equations than the standard one based on the 
subgroup $SO(3)$. Our calculation elucidated the explicit role of the local $SO(3)$ and 
$SO(3,1)$ invariance as well as translation invariance. In particular, in the $so(3)$ theory, 
the $J_3$ generator commutes with the generator of translations in the $z$-direction, a
property that is crucial for simplifying the equations. On the contrary, in the $su(2)\times su(2)$ 
representation, no $so(3,1)$ generator commutes with any of the translation generators. 
We speculate that 
this is the reason why the larger group, which most naturally matches the local 
symmetries of the metric tensor and the conformal form of the zeroth order cosmological solution, 
did not provide a simpler system of equations than the standard formulation.

Our calculation provides a blue-print for decomposing tensor quantities using group theoretical  
considerations in different perturbation theories of Einstein's field equations. Possible 
generalizations of the group theoretical formulation for cosmological perturbation theory 
includes different representations of the Lorentz (or the conformal Poincar\'e) group, using 
for example the full conformal group generators for decomposition, fully spinorial perturbation 
theory, or generalization to higher dimensions using the corresponding local symmetries of the 
metric tensor. The presented group theoretical approach could also be useful beyond the linear order, 
or considering inhomogeneous cosmological models. Furthermore the formalism may be applied to 
perturbation theory of other, not necessarily cosmological, solutions of the Einstein equations
with different background symmetries.

\ack
Many parts of the calculations were performed and checked using the
computer algebra programs MATHEMATICA 7, MAPLE 13 and GRTensorII. 
I.S.~acknowledges support from NASA grants NNG06GE71G and NNX10AD53G, 
and from the Pol\'anyi Program of the Hungarian National Office for 
Research and Technology (NKTH). V.G.Cz.~was supported by the 
National Research Foundation of South Africa and the Hungarian National 
Research Fund, OTKA No.~K67790 grant.

\end{document}